\documentstyle[aps,graphicx]{revtex}
\def \D {\mbox{D}}
\def \curl {\mbox{curl}\,}

\def \cu{ \rho^*}
\def \cq{ q^*}
\def \cp{ \pi^*}
\def \bcq { \bar{q}^*}
\def \bcp { \bar{\pi}^*}
\def\be{\begin{equation}}
\def\ee{\end{equation}}
\def\bea{\begin{eqnarray}}
\def\eea{\end{eqnarray}}

\begin{document}

\title{BRANE-WORLD COSMOLOGICAL PERTURBATIONS
\\ A covariant approach }
\author{Roy Maartens}
\address{Institute of Cosmology \& Gravitation, Portsmouth University,
Portsmouth~PO1~2EG, Britain}

\maketitle

\begin{abstract}

The standard cosmological model, based on general relativity with
an inflationary era, is very effective in accounting for a broad
range of observed features of the universe. However, the ongoing
puzzles about the nature of dark matter and dark energy, together
with the problem of a fundamental theoretical framework for
inflation, indicate that cosmology may be probing the limits of
validity of general relativity. The early universe provides a
testing ground for theories of gravity, since gravitational
dynamics can lead to characteristic imprints on the CMB and other
cosmological observations. Precision cosmology is in principle a
means to constrain and possibly falsify candidate quantum gravity
theories like M~theory. Generalized Randall-Sundrum brane-worlds
provide a phenomenological means to test aspects of M~theory. I
outline the 1+3-covariant approach to cosmological perturbations
in these brane-worlds, and its application to CMB anisotropies.

\end{abstract}

\section{Introduction}
%%%%%%%%%%%%%%%%%%%%%%

M~theory, the 11-dimensional theory that encompasses the known
superstring theories, is only partially understood, but is widely
considered to be a promising potential route to quantum
gravity~\cite{gen}, and is therefore an important candidate for
cosmological testing. Currently there are not realistic M~theory
cosmological solutions, so that it is reasonable to use simplified
phenomenological models that share some of the key features of
M~theory, especially branes. In brane cosmology, the observable
universe is a 1+3-dimensional ``brane" surface moving in a
higher-dimensional ``bulk" spacetime. Fields and particles in the
non-gravitational sector are confined to the brane, while gravity
propagates in the bulk. The simplest, and yet sufficiently
general, phenomenological brane-world models are the cosmological
generalizations~\cite{bdel,sms,mwbh} of the Randall-Sundrum~II
model~\cite{rs}. In the RSII brane-world, the bulk is
5-dimensional anti-de Sitter spacetime, so that the extra
dimension is infinite. The generalized brane-worlds
(see~\cite{m2,rl} for recent reviews) also have non-compact extra
dimension.\footnote{%
Various extensions of the generalized RSII brane-worlds are not
discussed here. The simplest extension is to introduce a second
brane, so that the extra dimension is compact (but much larger
than Planck scale), or a scalar field (gravitational sector) in
the bulk, or both. See~\cite{rl} for further discussion and
references (including the ``ekpyrotic" and cyclic models where the
potential of the bulk scalar causes branes to collide, which may
initiate a big bang and provide an alternative to
inflation~\cite{ec}). Other extensions involve corrections to the
action, including 4-dimensional ``induced gravity" and
5-dimensional Gauss-Bonnet corrections; see, e.g.~\cite{ext}. }
 The other 6 extra spatial dimensions of M~theory may be assumed to
be stabilized and compactified on a very small scale, so that they
do not affect the dynamics over the range of validity of the
brane-world model, i.e. for energies sufficiently below the string
scale, for which the brane may be treated as infinitely thin. The
RSII models have the additional advantage that they provide a
framework for investigating aspects of holography and the AdS/CFT
correspondence.

What prevents gravity from `leaking' into the infinite extra
dimension at low energies is the negative bulk cosmological
constant,
 \be
 \Lambda_5=-{6\over \ell^2},
 \ee
where $\ell$ is the curvature radius if the bulk is AdS$_5$.
Corrections to Newton's law in the weak-field static limit are
$O(\ell^2/r^2)$~\cite{rs}:
 \be\label{newt}
\phi(r) = {GM\over r}\left(1+{2\ell^2\over 3r^2}\right)+\cdots
 \ee
Experiments currently impose an upper bound $\ell \lesssim 1$~mm.
On the brane, the negative $\Lambda_5$ is offset by the positive
brane tension $\lambda$.  The effective cosmological constant on
the brane is
\begin{equation}\label{cc}
\Lambda=\frac{1}{2} (\Lambda_5+\kappa^2\lambda)\,,~~\kappa^2= 8\pi
G={8\pi\over M_4^2}\,,
\end{equation}
where $M_4\sim 10^{19}~$GeV is the effective Planck scale on the
brane. This is not the true fundamental gravity scale, which can
be much lower, offering the possibility of a resolution of the
hierarchy problem, as well as the exciting prospect that quantum
gravity effects could be observable in particle accelerators and
cosmic ray showers. The fundamental energy scale can be as low as
$\sim$~TeV in some brane-world scenarios, but in generalized RSII
models it is higher, $M_5>10^5~$TeV, and is related to $M_4$ via
\begin{equation}\label{scales}
M_5^3=\frac{M_4^2}{\ell}\,.
\end{equation}
The bound $\ell<1~$mm implies that $\lambda$ is above the
electroweak scale, $\lambda>(100~{\rm GeV})^4$. At high energies
($\rho\gg\lambda$) in the early universe, gravity becomes
5-dimensional and there are significant corrections to standard
cosmological dynamics. There are also corrections that can operate
at low energies, mediated by bulk graviton or Kaluza-Klein (KK)
modes. Both types of correction play an important role in
cosmological perturbations. In particular, 5-dimensional
gravitational-wave modes introduce nonlocal effects from the
viewpoint of brane-bound observers~\cite{mu,m1}.

\subsection*{Brane-world inflation}
%%%%%%%%%%%%%%%%%%%%%%%%%%%%%%%%%%%

The unperturbed cosmological brane-world is a Friedmann brane in a
Schwarzschild-AdS$_5$ bulk~\cite{bdel,unpert}. High-energy
brane-world modifications to the dynamics of inflation on the
brane have been investigated~\cite{mwbh,inf}. Essentially, the
high-energy corrections provide increased Hubble damping,
 \be
V(\varphi)\gg\lambda~\Rightarrow~ H\approx {V\over
M_{4}\sqrt{6\lambda}}\,,
 \ee
thus making slow-roll inflation possible even for potentials
$V(\varphi)$ that would be too steep in standard
cosmology~\cite{mwbh,steep}. This can be seen clearly from the
slow-roll parameters ($V\gg\lambda$)
 \bea
\epsilon\approx\epsilon_{\rm gr}\left({ {4\lambda\over
V}}\right)\,,~ \eta\approx\eta_{\rm gr}\left({ {2\lambda\over
V}}\right),
 \eea
where $\epsilon_{\rm gr},\eta_{\rm gr}$ are the standard general
relativity slow-roll parameters. Steep potentials can inflate at
high energy and then naturally stop inflating when $V$ drops below
$\lambda$. These models can be constrained because they typically
generate a blue spectrum of gravitational waves which can disturb
nucleosynthesis~\cite{steep}. They also raise the intriguing
prospect that the inflaton could act as dark matter or
quintessence at low energies~\cite{steep,dq}.

Large-scale scalar perturbations generated by slow-roll inflation
($V\gg\lambda$) have an enhanced amplitude compared with the
standard general relativity case~\cite{mwbh}:
 \be
A_{\rm s}^2 \approx  \left[{64\pi\over 75 M_{4}^6}\,{V^3\over
V'^2}\right]\left( { {V\over\lambda }}\right)^2.
 \ee
This means that COBE-scale perturbations can be generated when the
inflaton is well below $M_4$. For example,
 \be
\varphi_{\rm cobe}\approx \frac{300}{(M_4\ell)^{1/3}}\,M_4~\ll
M_4\,,
 \ee
for $V={1\over2}m^2\varphi^2$. The scalar spectral index is in
general given by
 \be
n_{\rm s}= 1-6\epsilon+2\eta\,,
 \ee
and is driven closer to 1 (compared to general relativity) by
high-energy effects.

High-energy inflation on the brane also generates a zero-mode
(4-dimensional graviton mode) of tensor perturbations, and
stretches it to super-Hubble scales. This zero-mode has the same
qualitative features as in general relativity, remaining frozen at
constant amplitude while beyond the Hubble horizon. Its amplitude
is enhanced at high energies, although the enhancement is much
less than for scalar perturbations~\cite{lmw}:
 \bea
A_{\rm \,t}^2 &\approx& \left[{8V\over 25 M_{4}^2}\right] \left(
{V \over \lambda}\right)^2,\\  {A_{\rm \,t}^2\over A_{\rm s}^2}
&\approx& \left[{3M_{4}^2\over8\pi}\,{V'^2\over V^2}\right] \left(
{\lambda\over V}\right).\label{ten}
 \eea
Equation~(\ref{ten}) means that brane-world effects suppress the
large-scale tensor contribution to CMB anisotropies. The tensor
spectral index has a smaller magnitude than in general relativity,
but obeys the same consistency relation:
 \be
n_{\rm t}=-3\epsilon = -2{A_{\rm \,t}^2\over A_{\rm s}^2}\,.
 \ee

The massive KK modes (5-dimensional graviton modes which have an
effective mass from a brane observer viewpoint) remain in the
vacuum state during slow-roll inflation~\cite{lmw,grs}. The
evolution of the super-Hubble zero mode is the same as in general
relativity, so that high-energy brane-world effects in the early
universe serve only to rescale the amplitude. However, when the
zero mode re-enters the Hubble horizon, massive KK modes can be
excited. This may be a very small effect, but it remains to be
properly quantified.

Vector perturbations in the bulk metric can support vector metric
perturbations on the brane, even in the absence of matter
perturbations. However, there is no normalizable zero mode, and
the massive KK modes stay in the vacuum state during brane-world
inflation~\cite{bmw}. Therefore, as in general relativity, we can
neglect vector perturbations in inflationary cosmology.

\subsection*{Perturbation evolution}

The background dynamics of brane-world cosmology is known exactly.
Large-scale cosmological perturbations on the brane are well
understood~\cite{mwbh,m1,lmw,gm,lmsw}. However, without a solution
for small-scale perturbations, we remain unable to predict the CMB
anisotropies in brane-world cosmology, and the CMB provides the
key means to test the scenario. The problem is that the
5-dimensional bulk perturbation equations must be solved in order
to solve for perturbations on the brane. The 5-dimensional
equations are partial differential equations for the 3-dimensional
Fourier modes, with complicated boundary conditions. In fact, even
the Sachs-Wolfe effect requires information from the 5-dimensional
solutions; although the large-scale density perturbations can be
determined without knowing the 5-dimensional
solutions~\cite{m1,gm}, the Sachs-Wolfe effect requires the
large-scale metric perturbations, and these are related to the
density perturbations in a way that involves the KK
modes~\cite{lmsw}.

The theory of gauge-invariant perturbations in brane-world
cosmology has been extensively investigated and
developed~\cite{mwbh,mu,m1,steep,lmw,grs,bmw,gm,lmsw,pert,bm,l1,l2}
and is qualitatively well understood. The key remaining task is
integration of the coupled brane-bulk perturbation equations; up
to now, only special cases have been solved, where these equations
effectively decouple. In general, and for the crucial case of
calculating CMB anisotropies~\cite{lmsw,bm,l1,l2}, the coupled
system must be solved. From the brane viewpoint, the bulk effects,
i.e. the high-energy corrections and the KK modes, act as source
terms for the brane perturbation equations. At the same time,
perturbations of matter on the brane can generate KK modes (i.e.,
emit 5-dimensional gravitons into the bulk) which propagate in the
bulk and can interact with the brane. This nonlocal interaction
amongst the perturbations is at the core of the complexity of the
problem. It can be elegantly expressed via integro-differential
equations~\cite{mu}, which take the form
 \be \label{ide}
A_k(t)=\int dt'\,{\cal G}(t,t') B_k(t')\,,
 \ee
where ${\cal G}$ is the bulk retarded Green's function evaluated
on the brane, and $A_k, B_k$ are made up of brane metric and
matter perturbations and their (brane) derivatives, and include
high-energy corrections to the background dynamics.

\section{Covariant dynamics and perturbations}
%%%%%%%%%%%%%%%%%%%%%%%%%%%%%%%%%%%%%%%%%%%%%%%%%%

The 5D field equations are
 \be
^{(5)}\!G_{AB}=-\Lambda_5\,
^{(5)}\!g_{AB}+\delta(y)\,\frac{8\pi}{M_5^3}\left[ -\lambda
g_{AB}+T_{AB}\right],
 \ee
where $y$ is a Gaussian normal coordinate orthogonal to the brane,
which is at $y=0$, the induced metric on $\{y=\mbox{ const}\}$ is
$g_{AB}= {}^{(5)}\!g_{AB}-n_An_B$ with $n^A$ the unit normal, and
$T_{AB}$ is the energy-momentum tensor of particles and fields
confined to the brane (with $T_{AB}n^B=0$). The effective field
equations on the brane are derived from the Gauss-Codazzi
equations and the Darmois-Israel junction conditions (using
$Z_2$-symmetry)~\cite{sms}:
\begin{equation} \label{e:einstein1}
G_{ab} = - \Lambda g_{ab} + \kappa^2 T_{ab} +
6\frac{\kappa^2}{\lambda} {\cal S}_{ab} - {\cal E}_{ab}\;,
\end{equation}
where ${\cal S}_{ab}\sim (T_{ab})^2$ is the high-energy correction
term, which is negligible for $\rho\ll\lambda$, while ${\cal
E}_{ab}$ is the projection of the bulk Weyl tensor on the brane.
This term encodes corrections from KK or 5D graviton effects. From
the brane-observer viewpoint, the energy-momentum corrections in
${\cal S}_{ab}$ are local, whereas the KK corrections in ${\cal
E}_{ab}$ are nonlocal, since they incorporate 5D gravity wave
modes, as discussed above. These nonlocal corrections cannot be
determined purely from data on the brane, and so the effective
field equations are not a closed system. One needs to supplement
them by 5D equations governing ${\cal E}_{ab}$, which are obtained
from the 5D Einstein and Bianchi equations~\cite{sms}.

The trace free ${\cal E}_{ab}$ contributes an effective energy
density $\rho^*$, pressure $\rho^*/3$, momentum density $q^*_a$
and anisotropic stress $\pi^*_{ab}$ on the brane. In a
1+3-covariant decomposition,
 \be
-{1\over\kappa^2} {\cal E}_{ab} = {\cu}\left(u_a u_b+{ {1\over3}}
h_{ab}\right)+ {\cq_a} u_{b} + {\cq_b} u_{a}+\cp_{ab}\,,
 \ee
where $u^a$ is a physically determined 4-velocity on the brane and
$h_{ab}=g_{ab}+u_au_b$ projects into the comoving rest space at
each event. The KK anisotropic stress $\cp_{ab}$ incorporates the
spin-0 (``Coulomb"), spin-1 (gravimagnetic) and spin-2
(gravitational wave) 4D modes of the 5D graviton. The KK momentum
density $\cq_a $ incorporates spin-0 and spin-1 modes, and the KK
energy density $\cu $ (the ``dark radiation") incorporates the
spin-0 mode. The brane ``feels" the bulk gravitational field
through these terms. In the background, $q^*_a=0=\pi^*_{ab}$,
since only the ``dark radiation" term is compatible with Friedmann
symmetry.

The brane-world corrections can conveniently be consolidated into
an effective total energy density, pressure, momentum density and
anisotropic stress. Linearizing the general nonlinear
expressions~\cite{m1} we obtain
\begin{eqnarray}
\rho^{\text{eff}} &=& \rho\left(1 +\frac{\rho}{2\lambda} +
\frac{\rho^*}{\rho} \right)\;, \\ \label{e:pressure1} p^{\text{eff
}} &=& p  + \frac{\rho}{2\lambda} (2p+\rho)+\frac{\rho^*}{3}\;, \\
q^{\text{eff }}_a &=& q_a\left(1+  \frac{\rho}{\lambda} \right)
+q^*_a\;, \\ \label{e:pressure2} \pi^{\text{eff }}_{ab} &=&
\pi_{ab} \left(1-\frac{\rho+3p}{2\lambda}\right)+\pi^*_{ab}\;,
\end{eqnarray}
where $\rho=\sum_i\rho^{(i)}$ and $p=\sum_ip^{(i)}$ are the total
matter-radiation density and pressure, $q_a=\sum_iq^{(i)}_a$ is
the total matter-radiation momentum density, and $\pi_{ab}$ is the
photon anisotropic stress (neglecting that of neutrinos, baryons
and CDM).

Energy-momentum conservation,
 \be \label{lc}
\nabla^b T_{ab}=0\,,
 \ee
together with the 4D Bianchi identity, lead to
 \be \label{nlc}
\nabla^a{\cal E}_{ab}={6\kappa^2\over\lambda}\,\nabla^a{\cal
S}_{ab}\,,
 \ee
which shows qualitatively how 1+3 spacetime variations in the
matter-radiation on the brane can source KK modes. The
1+3-covariant decomposition of Eq.~(\ref{lc}) leads to the
standard energy and (linearized) momentum conservation equations,
\begin{eqnarray}
&&\dot{\rho}+\Theta(\rho+p)+\D^aq_a=0\,,\\ && \dot{q}_a+ 4Hq_a+
\D_a p+(\rho+p)A_a+ \D^b\pi_{ab} =0\,,\label{c2}
\end{eqnarray}
where $\Theta$ is the volume expansion rate, which reduces to $3H$
in the background ($H$ is the background Hubble rate), $A_a$ is
the 4-acceleration, and $\D_a$ is the covariant derivative in the
rest space (i.e. $\D_aF^{b\cdots}{}{}{}_{\cdots
c}=h_a{}^dh^b{}_e\cdots h_c{}^f \nabla_d F^{e\cdots}{}{}
{}_{\cdots f}$). The absence of bulk source terms in the
conservation equations is a consequence of having $\Lambda_5$ as
the only 5D source in the bulk. If there is a bulk scalar field,
then there is energy-momentum exchange between the brane and bulk
(in addition to the gravitational interaction)~\cite{sca}.

Equation~(\ref{nlc}) may be thought of as the ``nonlocal
conservation equation". Linearizing the general 1+3-covariant
decomposition~\cite{m1}, we obtain
\begin{eqnarray}
&& \dot{\rho}^*+{{4\over3}}\Theta{\cu}+\D^a{\cq_a}=0\,,
\label{nlc1}
\\&& \dot{q}^*_a+4H{\cq_a}
+{{1\over3}}\D_a{\cu}+{{4\over3}}{\cu}A_a +\D^b{\cp_{ab}} ={
(\rho+p)\over\lambda}\left[-\D_a \rho +3Hq_a+{3\over2}
\D^b\pi_{ab} \right].\label{nlc2}
\end{eqnarray}
At linear order, spatial inhomogeneity ($\D_a\rho$), peculiar
motions ($q_a=\rho v_a$) and anisotropic stresses ($\pi_{ab}$) in
the matter-radiation on the brane are seen to be sources for KK
modes (or 5D graviton emission into the bulk). Qualitatively and
geometrically this can be understood as follows: the non-uniform
5D gravitational field generated by inhomogeneous and anisotropic
4D matter-radiation contributes to the 5D Weyl tensor, which
nonlocally ``backreacts" on the brane via its projection ${\cal
E}_{ab}$. Note also that the source terms are suppressed at low
energies, and during quasi-de Sitter inflation on the brane.

Equations~(\ref{nlc1}) and (\ref{nlc2}) are propagation equations
for $\cu$ and $\cq_a$. There is no propagation equation on the
brane for $\cp_{ab}$; if there were such an equation, then one
could determine the KK modes purely from data on the brane, which
would violate causality for 5D gravitational waves.

In the background, the modified Friedmann equations are
\begin{eqnarray}
H^2 &=& \frac{\kappa^2}{3} \rho^{\text{eff }} + \frac{1}{3}
\Lambda + \frac{K}{a^2} \,, \\ \label{e:friedmann1} \dot H &=&
-\frac{\kappa^2}{2}(\rho^{\rm eff} +p^{\rm eff})+\frac{K}{a^2}\,.
\end{eqnarray}
By Eq.~(\ref{nlc1}), the KK energy density behaves like dark
radiation:
\begin{equation}
\rho^* \propto\frac{1}{a^4}\,.
\end{equation}
The source of the dark radiation is the tidal (Coulomb) effect of
a 5D black hole in the bulk. When the black hole mass vanishes,
the bulk geometry reduces to AdS$_5$ and $\cu=0$. In order to
avoid a naked singularity, we assume that the black hole mass is
non-negative, so that $\cu\geq0$. This additional effective
relativistic degree of freedom is constrained by nucleosynthesis
and CMB observations to be no more than $\sim$3\% of the radiation
energy density~\cite{lmsw,bm,dr}:
 \be
\left. {\cu \over \rho_{\rm rad}}\right|_{\rm nuc} \lesssim 0.03
 \ee

If $\cu=0$ and $K=0=\Lambda$, then the exact solution of the
Friedmann equations is~\cite{bdel}
 \be\label{ex1}
a=\,{\rm const}\,[t(t+t_{\lambda})]^{1/3(w+1)}\,, ~~
t_{\lambda}={M_{4}\over\sqrt{\pi\lambda}}< 10^{-9}\, {\rm sec}\,,
 \ee
where $w=p/\rho$ is assumed constant. If $\cu\neq0$ (but
$K=0=\Lambda$), then the solution for the radiation era
($w={1\over3}$) is~\cite{bm}
 \be\label{ex2}
a=\,{\rm const}\,[t(t+t_{\lambda})]^{1/4}\,, ~~
t_{\lambda}={\sqrt{3}\,M_{4}\over
4\sqrt{\pi\lambda}\,(1+\cu/\rho)}\,.
 \ee
For $t\gg t_\lambda$ we recover from Eqs.~(\ref{ex1}) and
(\ref{ex2}) the standard behaviour, $a\propto t^{2/3(w+1)}$,
whereas for $t\ll t_\lambda$, we have the very different
behaviour, $a\propto t^{1/3(w+1)}$.

In the 1+3-covariant description of perturbations~\cite{m1}, we
isolate the KK anisotropic stress $\cp_{ab}$ as the term that must
be determined from 5D equations. Once $\cp_{ab}$ is determined in
this way, the 1+3 perturbation equations on the brane form a
closed system. The KK terms act as source terms modifying the
standard general relativity perturbation equations, together with
the high-energy corrections. For example, the propagation equation
for the shear is~\cite{m1}
 \be
\dot{\sigma}_{ab}+2H\sigma_{ab}+ E_{ab}-{\kappa^2\over2} \pi_{ab}-
\D_{\langle a}A_{b\rangle} = {\kappa^2\over 2}\cp_{ab}-
{\kappa^2\over 4}(1+3w){\rho\over\lambda}\,\pi_{ab}\,,
 \ee
where $E_{ab}$ is the electric part of the 4D brane Weyl tensor
(not to be confused with ${\cal E}_{ab}$). In general relativity,
the right hand side is zero. In the brane-world, the first source
term on the right is the KK term, the second term is the
high-energy modification. The other modification is a
straightforward high-energy correction of the background
quantities $H$ and $\rho$ via the modified Friedmann equations.

In the 1+3-covariant approach, perturbative quantities are
projected vectors ($V_au^a=0$) and projected symmetric tracefree
tensors,
 \be
W_{ab}=W_{\langle ab\rangle}\equiv
\left[h_a{}^ch_b{}^d-{1\over3}h_{ab}h^{cd}\right]W_{cd}\,.
 \ee
These are decomposed into (3D) scalar, vector and tensor modes
as~\cite{m2}
\begin{eqnarray}
 V_a &=& \D_a V+\bar{V}_a\,, \\
 W_{ab} &=& \D_{\langle a}\D_{b\rangle}{W}
+\D_{\langle a}\bar{W}_{b\rangle}+\bar{W}_{ab}\,,
\end{eqnarray}
where an overbar denotes a (3D) transverse quantity
($\D^a\bar{V}_a=0= \D^b\bar{W}_{ab}$). Purely scalar perturbations
are characterized by
 \be
\bar{V}_a=\bar{W}_a=\bar{W}_{ab}=0\,,
 \ee
and scalar quantities are formed via the (3D) Laplacian: ${\cal
V}=\D^a\D_a V\equiv \D^2 V$. Purely vector perturbations are
characterized by
 \be\label{vec}
V_a=\bar{V}_a\,,~W_{ab}=\D_{\langle a}
\bar{W}_{b\rangle}\,,~\curl\D_a f=-2\dot{f}\omega_a\,,
 \ee
where $\omega_a$ is the vorticity, and purely tensor by
 \be
\D_a f=0=V_a\,,~ W_{ab}=\bar{W}_{ab}\,.
 \ee

The KK energy density produces a scalar mode $\D_a{\cu}$ (which is
present even if $\cu=0$ in the background). The KK momentum
density carries scalar and vector modes, and the KK anisotropic
stress carries scalar, vector and tensor modes:
\begin{eqnarray}
{\cq_a}&=&\D_a{\cq}+{\bcq_a}\,,\\ {\cp_{ab}}&=&\D_{\langle
a}\D_{b\rangle}{\cp} +\D_{\langle
a}{\bcp_{b\rangle}}+{\bcp_{ab}}\,.
\end{eqnarray}

\subsection*{ Density perturbations}
%%%%%%%%%%%%%%%%%%%%%%%%%%%%%%%%%%%%

We define matter density and expansion (velocity) perturbation
scalars, as in general relativity,
 \be
\Delta={a^2\over\rho}\D^2\rho\,,~Z=a^2\D^2\Theta\,,
 \ee
and then define dimensionless KK scalars~\cite{m1},
 \be
{   U}={a^2\over\rho}\D^2{\cu}\,,~{   Q}={a\over\rho} \D^2
{\cq}\,,~{  \Pi }={1\over \rho}\D^2{\cp}\,.
 \ee
A non-adiabatic isocurvature mode is associated with the KK
fluctuations. We can see this from the 1+3-covariant expression
which determines the non-adiabatic perturbations in the matter
plus KK ``fluid"~\cite{gm}:
\begin{eqnarray}
\dot{\rho}^{\rm \,eff}\,\D^2 p^{\rm eff}-\dot{p}^{\rm \,eff}\,\D^2
\rho^{\rm eff}={H\rho\over 9a^2} \left[c_{\rm s}^2 - {1\over 3}
+\left({2\over 3}+w+c_{\rm s}^2\right){ {\rho\over\lambda}}\right]
\left( 3\rho { U}- 4{\cu}\Delta \right)\,,
\end{eqnarray}
where $c_{\rm s}^2=\dot{p}/\dot{\rho}$, and the matter
perturbations are assumed adiabatic. This mode is in general
present, in particular when $\cu=0$ in the background. However the
mode will in general decay and be suppressed at low energies.

This can be related to the curvature perturbation ${\cal R}$ on
uniform density surfaces, which is defined in the metric-based
perturbation theory. The associated gauge-invariant quantity
 \be
\xi={\cal R}+ {\delta\rho \over 3(\rho+p)}\,,
 \ee
may be defined for matter on the brane ($\xi^{\rm m}$), for the KK
``fluid" ($\xi^*$) if $\cu\neq0$, and for the total, effective
fluid ($\xi^{\rm eff}$). If $\cu\neq0$ in the background,
 \be
{\xi}^{\rm eff}= {\xi}^{\rm m}+\left[{4\cu \over 3(\rho+ p)(1+
\rho/\lambda)+4\cu}\right] \left(\xi^*-\xi^{\rm m}\right)\,.
 \ee
In the case where $\cu=0$ in the background, the evolution of the
total curvature perturbation on large scales is~\cite{lmsw}:
 \be
\dot{\xi}^{\rm \,eff}= \dot{\xi}^{\rm \,m}+H\left[c_{\rm
s}^2-{1\over 3}+\left({\rho+p \over \rho+ \lambda}\right)\right]
{\delta\cu \over (\rho+p)(1+\rho/\lambda)}\,.
 \ee
For adiabatic matter perturbations, $\dot{\xi}^{\rm \,m}=0$,
independent of brane-world modifications to the field equations,
since this result depends on energy conservation
only~\cite{encon}. However, $\dot{\xi}^{\rm \,eff}\neq 0$ even for
adiabatic matter perturbations. The KK effects on the brane
contribute a non-adiabatic mode, although $\dot{\xi}^{\rm
\,eff}\to 0$ at low energies.

\begin{figure}[!bth]
\begin{center}
\includegraphics[scale=1.2]{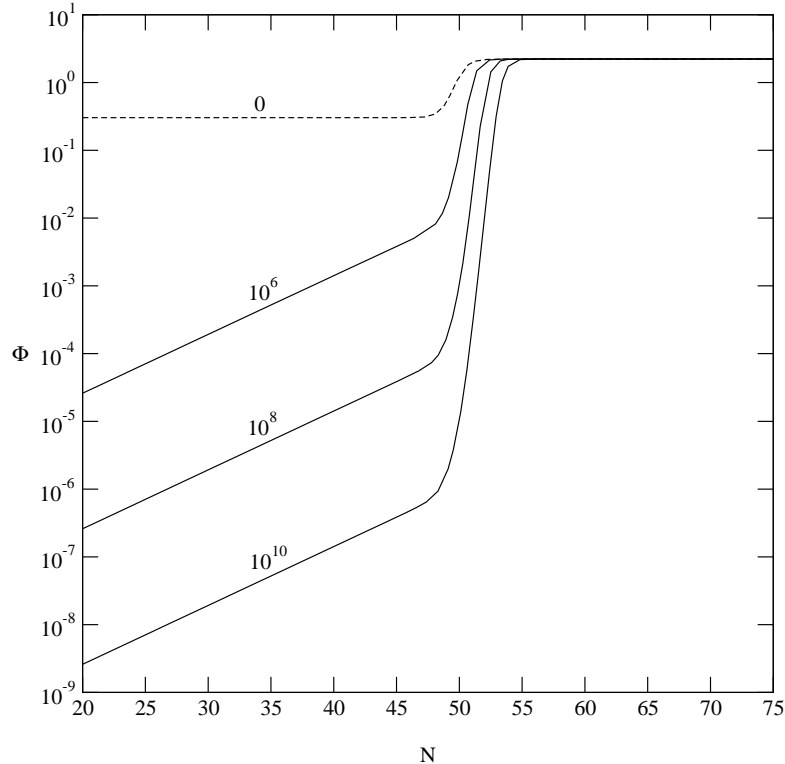}
\caption{The evolution of $\Phi$ along a fundamental world-line
for a mode that is well beyond the Hubble horizon at
$N\equiv\ln(a/a_0) =0$, about 50 e-folds before inflation ends,
and remains super-Hubble through the radiation era. A smooth
transition from inflation to radiation is modelled by
$w={1\over3}[(2-\alpha)\tanh(N-50)-(1-\alpha)]$, where $\alpha$ is
a small positive parameter (chosen as $\alpha=0.1$ in the plot).
Labels on the curves indicate the value of $\rho_0/\lambda$, so
that the general relativistic solution is the dashed curve
($\rho_0/\lambda=0$). For $\rho_0/\lambda\gg1$, inflation ends at
$N=50-2\ln[(1-2\alpha)/3]\approx 47.4$, and at $N=50$ in general
relativity. Only the lowest curve still has $\rho/\lambda\gg1$ at
the start of radiation-domination ($N$ greater than about 53), and
one can see that $\Phi$ is still growing. }
\end{center}
\end{figure}

The covariant density perturbation equations on the brane reduce
to~\cite{gm}
\begin{eqnarray}
\dot{\Delta} &=&3wH\Delta-(1+w)Z\,, \\  \dot{Z}
&=&-2HZ-\left({c_{\rm s}^2\over 1+w}\right)
\D^2\Delta-\kappa^2\rho {   U}-{{1\over2}}\kappa^2 \rho\left[1+
(4+3w){ {\rho\over\lambda}}- \left({4c_{\rm s}^2\over
1+w}\right){\cu\over\rho}\right] \Delta \,,\\  {   \dot{U}} &=&
(3w-1)H{   U} + \left({4c_{\rm s}^2\over 1+w}\right)\left({{\cu
}\over\rho}\right) H\Delta -\left({4{\cu }\over3\rho}\right)
Z-a\D^2{ Q}\,,\\  {   \dot{Q}} &=&(3-1w)H{ Q}-{1\over3a}{
U}-{{2\over3}} a{ \D^2\Pi}+{1\over3 a}\left[ \left({4c_{\rm
s}^2\over 1+w}\right){{\cu}\over\rho}-3(1+w) {
{\rho\over\lambda}}\right]\Delta\,.
\end{eqnarray}
It follows that the system closes on super-Hubble scales, since
the KK anisotropic stress term $\Pi$ occurs only via its
Laplacian~\cite{m1}. KK effects introduce two new isocurvature
modes on large scales (associated with $U$ and $Q$), as well as
modifying the evolution of the adiabatic modes~\cite{gm,l1}. A
simple illustration of this modification is shown in Fig.~1
(from~\cite{gm}). The variable
 \be
\Phi=\kappa^2a^2\rho \Delta
 \ee
is a covariant analogue of the Bardeen metric perturbation
variable $\Phi_H$. The large-scale behaviour of $\Phi$ through the
inflationary and radiation eras is compared for different values
of the inflationary energy scale relative to the brane tension. In
general relativity, $\Phi$ is constant in both eras, whereas
$\Phi$ grows during the high-energy regime on the brane.

Although the density perturbations can be found on super-Hubble
scales, the Sachs-Wolfe effect requires $\Pi$ in order to
translate from density to metric perturbations. In the
longitudinal gauge of the metric perturbation formalism, the
gauge-invariant metric perturbations at last scattering are
related by
 \be
\Phi_A-\Phi_H= -\kappa^2a^2\delta \cp\,,
 \ee
where the radiation anisotropic stress on large scales is
neglected, as in general relativity, and $\delta\cp$ is equivalent
to the covariant quantity $\Pi$. In general relativity, the right
hand side is zero. The brane-world corrections to the general
relativistic direct Sachs-Wolfe effect are given by~\cite{lmsw}
 \be
{\delta T\over T} = \left({\delta T\over T}\right)_{\rm gr}
-{8\over 3}\left({\rho_{\rm rad}\over \rho_{\rm
cdm}}\right)S^*-\kappa^2a^2\delta \cp +{2\kappa^2\over
a^{5/2}}\int da\,\, a^{7/2}\,\delta\cp \,,
 \ee
where $S^*$ is the KK entropy perturbation (determined by
$\delta\cu$). The KK term $\delta\cp$ cannot be determined by the
4D brane equations, so that $\delta T/T$ cannot be evaluated on
large scales without solving the 5D equations.

A simple approximation to $\delta\cp$ on large scales is discussed
in~\cite{bm} and the Sachs-Wolfe effect is estimated as
 \be
{\delta T\over T}\sim \left({\delta\cp\over\rho}\right)_{\rm in}
\left({t_{\rm eq}\over t_{\rm ls}}\right)^{2/3} \left[{\ln(t_{\rm
in}/t_4)\over \ln (t_{\rm eq}/t_4)} \right],
 \ee
where $t_4$ is the 4D Planck time and $t_{\rm in}$ is the time
when the KK anisotropic stress is induced on the brane, which is
expected to be of the order of the 5D Planck time.

\subsection*{ Vector perturbations}
%%%%%%%%%%%%%%%%%%%%%%%%%%%%%%%%%%%

The vorticity propagation equation on the brane is the same as in
general relativity,
 \be\label{vor}
\dot{\omega}_a+2H\omega_a=-{1\over2}\curl A_a\,.
 \ee
Taking the curl of the conservation equation~(\ref{c2}) (for the
case of a perfect fluid, $q_a=0=\pi_{ab}$), and using the identity
in Eq.~(\ref{vec}), one obtains
 \be
\curl A_a=-6Hc_{\rm s}^2\omega_a\,,
 \ee
as in general relativity, so that Eq.~(\ref{vor}) becomes
 \be
\dot{\omega}_a+[2-3c_{\rm s}^2]H\omega_a=0\,,
 \ee
which expresses the conservation of angular momentum. In general
relativity, vector perturbations vanish when the vorticity is
zero. By contrast, in brane-world cosmology, bulk effects can
source vector perturbations even in the absence of
vorticity~\cite{m2}. This can be seen via the divergence equation
for the magnetic part $H_{ab}$ of the 4D Weyl tensor on the brane:
\begin{eqnarray}
\D^2\bar{H}_{a} = 2\kappa^2(\rho+p)\left[1+{
{\rho\over\lambda}}\right] \omega_a + {4\over3} \kappa^2{\cu}
\omega_a - {1\over2}\kappa^2 \curl{\bcq_a}\,,
\end{eqnarray}
where $H_{ab}=\D_{\langle a}\bar{H}_{b\rangle}$. Even when
$\omega_a=0$, there is a source for gravimagnetic terms on the
brane from the KK quantity $\curl{\bcq_a}$.

We define covariant dimensionless vector perturbation quantities
for the vorticity and the KK gravimagnetic term:
 \be
\bar{\alpha}_a=a\,\omega_a\,,~
\bar{\beta}_a={a\over\rho}\curl{\bcq_a}\,.
 \ee
On large scales, we can find a closed system for these vector
perturbations on the brane~\cite{m2}:
\begin{eqnarray}
\dot{\bar{\alpha}}_a+\left(1-3c_{\rm s}^2\right)
H\bar{\alpha}_a&=&0 \,,\\ \dot{\bar{\beta}}_a+(1-3w)H\bar{\beta}_a
&=& {2\over 3}H\left[4\left( 3c_{\rm s}^2-1\right)
{{\cu}\over\rho}- 9(1+w)^2{
{\rho\over\lambda}}\right]\bar{\alpha}_a\,.\label{v}
\end{eqnarray}
Thus we can solve for $\bar{\alpha}_a$ and $\bar{\beta}_a$ on
super-Hubble scales, as for density perturbations. Vorticity in
the brane matter is a source for the KK vector perturbation
$\bar{\beta}_a$ on large scales. Vorticity decays unless the
matter is ultra-relativistic or stiffer ($w\geq {1\over3}$), and
this source term typically provides a decaying mode. There is
another pure KK mode, independent of vorticity, but this mode
decays like vorticity. For $w\equiv p/\rho=\,$const, the solutions
are
 \bea
\bar{\alpha}_a &=& b_a \left({a\over a_0}\right)^{3w-1}\,,\\
\bar{\beta}_a &=& c_a\left({a\over a_0}\right)^{3w-1}+ b_a\left[
\epsilon_w\,{8\cu_0 \over 3 \rho_0}\left({a\over
a_0}\right)^{2(3w-1)}+2(1+w){\rho_0\over \lambda} \left({a\over
a_0}\right)^{-4}\right]\,,
 \eea
where $\dot{b}_a=0=\dot{c}_a$ and $\epsilon_w=(1,0)$ for $(w \neq
{1\over3},w = {1\over3})$.

Inflation will redshift away the vorticity  and the KK mode, which
is consistent with the analysis in~\cite{bmw} of vector
perturbations generated during inflation.

\subsection*{ Tensor perturbations}
%%%%%%%%%%%%%%%%%%%%%%%%%%%%%%%%%%%

The covariant description of tensor modes on the brane is via the
shear, which satisfies the wave equation~\cite{m2}
\begin{eqnarray}
&& \D^2{\bar{\sigma}}_{ab}-\ddot{\bar{\sigma}}_{ab}
-5H\dot{\bar{\sigma}}_{ab}-\left[2\Lambda+{{1\over2}}
\kappa^2\left\{\rho-3p- (\rho+3p){ {\rho\over\lambda}}\right\}
\right]{\bar{\sigma}}_{ab}\nonumber\\ &&~~{}=- {\kappa^2}\left( {
\dot{\bar{\pi}}^*_{ ab }}+2H {\bcp_{ab}} \right).
\end{eqnarray}
Unlike the density and vector perturbations, there is no closed
system on the brane for large scales. The KK anisotropic stress
$\bcp_{ab}$ is an unavoidable source for tensor modes on the
brane. These modes and their effect on the CMB are discussed in
the following section.

\section{CMB anisotropies in the brane-world}
%%%%%%%%%%%%%%%%%%%%%%%%%%%%%%%%%%%%%%%%%%%%%

The perturbation equations in the previous section form the basis
for an analysis of scalar and tensor CMB anisotropies in the
brane-world. The full system of equations on the brane, including
the Boltzmann equation for photons, has been given for
scalar~\cite{l1} and tensor~\cite{l2} perturbations. But the
systems are not closed, as discussed above, because of the
presence of the KK anisotropic stress  $\cp_{ab}$, which acts a
source term. For example, in the tight-coupling radiation era, the
scalar perturbation equations may be decoupled to give an equation
for the gravitational potential $\Phi$, defined by
 \be
E_{ab}=\D_{\langle a}\D_{b\rangle}\Phi\,.
 \ee
In general relativity, this equation in $\Phi$ has no source term,
but in the brane-world there is a source term made up of
$\cp_{ab}$ and its time-derivatives. At low energies
($\rho\ll\lambda$), and for a flat background ($K=0$), the
equation is~\cite{l1}
 \bea
&& 3x\Phi_k''+12\Phi_k'+x\Phi_k \nonumber\\ &&~~~{}={\mbox{const}
\over \lambda}\,\left[ \pi_k^{*\prime\prime}-{1\over x}\,
{\pi_k^{*\prime}} +\left({2\over x^3}- {3 \over x^2}+ {1\over x}
\right) \cp_k\right],
 \eea
where $x=k/(aH)$, a prime denotes $d/dx$, and $\Phi_k$, $\cp_k$
are the Fourier modes of $\Phi$ and $\cp_{ab}$. In general
relativity the right hand side is zero, so that the equation may
be solved for $\Phi_k$, and then for the remaining perturbative
variables, which gives the basis for initializing CMB numerical
integrations. At high energies, earlier in the radiation era, the
decoupled equation is fourth order~\cite{l1}:
 \bea
&& 729 x^2\Phi_k''''+3888x\Phi_k'''+(1782+54x^2) \Phi_k'' +144x
 \Phi_k'+(90+x^2)\Phi_k ={\mbox{const}
}\,\left[243\left(\! {\cp_k\over\rho}\!
\right)^{\!\prime\prime\prime\prime} +\right. \nonumber\\ &&~~~{}
\left.-{ 810 \over x}
\left(\!{\cp_k\over\rho}\!\right)^{\!\prime\prime\prime} +
{18(135+2x^2) \over x^2}
\left(\!{\cp_k\over\rho}\!\right)^{\!\prime\prime}-
{30(162+x^2)\over x^3}
\left(\!{\cp_k\over\rho}\!\right)^{\!\prime} +
{x^4+30(162+x^2)\over x^4}
\left(\!{\cp_k\over\rho}\!\right)\right].
 \eea

The formalism and machinery are ready to compute the temperature
and polarization anisotropies in brane-world cosmology, once a
solution, or at least an approximation, is given for $\cp_{ab}$.
The resulting power spectra will reveal the nature of the
brane-world imprint on CMB anisotropies, and would in principle
provide a means of constraining or possibly falsifying the
brane-world models. Once this is achieved, the implications for
the fundamental underlying theory, i.e. M~theory, would need to be
explored.

However, the first step required is the solution or estimate of
$\cp_{ab}$. This solution will be of the  form, expressed in
Fourier modes (and assuming no incoming 5D gravitational waves):
\begin{equation}\label{e:soln}
\pi^*_k(t) \propto \int d t'\,\,{\cal G}(t,t') F_k(t') \,,
\end{equation}
where ${\cal G}$ is a retarded Green's function evaluated on the
brane. The functional $F_k$ will be determined by the covariant
brane perturbation quantities and their derivatives. It is known
in the case of a Minkowski background~\cite{ssm}, but not in the
cosmological case. Once ${\cal G}$ and $F_k$ are determined or
estimated, the numerical integration in Eq.~(\ref{e:soln}) can in
principle be incorporated into a modified version of a CMB
numerical code.

In order to make some progress towards understanding brane-world
signatures on CMB anisotropies, we need to consider approximations
to the solution. The nonlocal nature of $\pi^*_k$, as reflected in
Eq.~(\ref{e:soln}), is fundamental, but is also the source of the
great complexity of the problem. The lowest level approximation to
$\pi^*_k$ is local. Despite removing the key aspect of the KK
anisotropic stress, we can get a feel for its influence on the CMB
if we capture at least part of its qualitative properties. The key
qualitative feature is that inhomogeneity and anisotropy on the
brane are a source for KK modes in the bulk which ``backreact" or
``feed back" onto the brane.

The simplest case is that of tensor perturbations. The transverse
traceless part of inhomogeneity and anisotropy on the brane is
given by the transverse traceless anisotropic stresses in the
geometry, i.e. by the photon anisotropic stress $\bar{\pi}_{ab}$
and the shear anisotropy $\bar{\sigma}_{ab}$. The photon
anisotropic stress in turn is sourced by the shear to lowest order
(neglecting the role of the octupole and higher Boltzmann
moments), so that $F_k\approx F[ \bar{\sigma}_k]$. The simplest
local approximation which reflects the essential qualitative
feature of the spin-2 KK modes is~\cite{l2}
\begin{equation} \label{e:Pansatz1}
\kappa^2 \bar{\pi}_{ab}^{*} = - \zeta H \bar{\sigma}_{ab}\,,
\end{equation}
where $\zeta$ is a dimensionless KK parameter, which is assumed to
be a comoving constant in a first approximation,
 \be
\dot\zeta=0\,.
 \ee
The limit $\zeta=0$ corresponds to no KK effects on the brane, and
$\zeta=0=\lambda^{-1}$ gives the general relativity limit.

For tensor perturbations, there is no freedom over the choice of
frame (i.e.\ $u^a$), and thus there is no gauge ambiguity in
Eq.~(\ref{e:Pansatz1}). However, for scalar (or vector)
perturbations, this relation could only hold in one frame, since
$\pi_{ab}^{*}$ is frame-invariant in linear theory while
$\sigma_{ab}$ is not:
 \be
u^a \to u^a+v^a~~ \Rightarrow~~ \cp_{ab}\to \cp_{ab}\,,~
\sigma_{ab}\to \sigma_{ab}+ \D_{\langle a}v_{b\rangle}\,.
 \ee
Thus for scalar perturbations, we would need an alternative,
frame-invariant, local approximation.

The approximation in Eq.~(\ref{e:Pansatz1}) has the qualitative
form of a shear viscosity, which suggests that KK effects lead to
a damping of tensor anisotropies. This is indeed consistent with
the conversion of part of the zero-mode at Hubble re-entry into
massive KK modes~\cite{lmw,grs}. The conversion may be understood
equivalently as the emission of KK gravitons into the bulk. This
leads to a loss of energy in the 4D graviton modes on the brane,
i.e. to an effective damping. The approximation in
Eq.~(\ref{e:Pansatz1}), although local, therefore also
incorporates this key feature qualitatively.

The 1+3-covariant transverse traceless quantities are the electric
($E_{ab}$) and magnetic parts of the brane Weyl tensor, the shear,
and the anisotropic stresses. They are expanded in electric
($Q_{ab}^{(k)} $) and magnetic ($\hat{Q}_{ab}^{(k)} $) parity
tensor harmonics~\cite{c}, with dimensionless coefficients:
\begin{eqnarray}
\bar{E}_{ab} &=& \sum_{k} \left(\frac{k}{a}\right)^2 \left[E_k
Q_{ab}^{(k)} + \hat{E}_k \hat{Q}_{ab}^{(k)} \right], \\
\bar{H}_{ab} &=& \sum_{k} \left(\frac{k}{a}\right)^2 \left[H_k
Q_{ab}^{(k)} + \hat{H}_k \hat{Q}_{ab}^{(k)} \right], \\
\bar{\sigma}_{ab} &=& \sum_{k} \frac{k}{a} \left[\sigma_k
Q_{ab}^{(k)} + \hat{\sigma}_k \hat{Q}_{ab}^{(k)} \right],  \\
\bar{\pi}_{ab} &=& \rho \sum_{k} \left[
\pi_{k} Q_{ab}^{(k)} + \hat{\pi}_k \hat{Q}_{ab}^{(k)} \right], \\
\bar{\pi}_{ab}^{*} &=& \rho \sum_{k} \left[ \pi_{k}^{*}
Q_{ab}^{(k)} + \hat{\pi}_k^{*} \hat{Q}_{ab}^{(k)} \right].
\end{eqnarray}
Using $\bar{H}_{ab}={\rm curl}\,\bar{\sigma}_{ab}$, we arrive at
the coupled equations~\cite{l2}
\begin{eqnarray} \label{e:sigmadot}
&& \frac{k}{a^2} \left(\sigma'_k + {\cal H} \sigma_k \right) +
\frac{k^2}{a^2} E_k - \frac{\kappa^2}{2} \rho \pi_k = -\kappa^2
(1+3w)\frac{\rho^2}{4\lambda}\pi_k  + \frac{\kappa^2}{2} \rho
\pi_k^{*}\,,\\ && \label{e:Edot} \frac{k^2}{a^2} \left(E'_k +
{\cal H} E_k \right) - k\left[\frac{k^2}{a^2} + \frac{3K}{a^2} -
 \frac{\kappa^2}{2} (1+w) \rho  \right] \sigma_k +
\frac{\kappa^2}{2} \rho \pi'_k -\frac{\kappa^2}{2} (3 w+2) {\cal
H}\rho \pi_k \nonumber\\ &&{}~= -\frac{\kappa^2}{12\lambda}
\left[6 k (1+w) \rho^2  \sigma_k - 3(\rho' + 3 p') \rho \pi_k - 3
(3w+1) \rho \left(\rho \pi'_k + \rho' \pi_k \right) - 9 (3w+1)
\rho^2 {\cal H} \pi_k \right] \nonumber\\ &&{}~~  - \frac{2}{3}
k\kappa^2\rho^* \sigma_k  - \frac{\kappa^2}{2} \left[ \rho
 \pi_k^{*\prime} +\left(\rho'+ {\cal H} \rho\right) \pi_k^{*} \right]\,,
\end{eqnarray}
where $\tau$ is conformal time, a prime denotes $d/d\tau$, ${\cal
H}=a'/a$, and the equation-of-state parameter $w$ is not assumed
constant. Equations~(\ref{e:sigmadot}) and (\ref{e:Edot}), with
all brane-world terms on the right-hand sides, determine the
tensor anisotropies in the CMB, once $\pi_k$ and $\pi^*_k$ are
given. The former is determined by the Boltzmann equation in the
usual way~\cite{c}, since high-energy corrections are negligible
at and after nucleosynthesis. The latter is given by the
approximation Eq.~(\ref{e:Pansatz1}), which gives
\begin{equation}\label{app}
\kappa^2 \rho \pi_{k}^{*} = - \zeta {\cal H} \frac{k}{a^2}
\sigma_k\,.
\end{equation}
We will also assume $K=0=\rho^*$ in the background. The KK
parameter $\zeta$ (together with the brane tension $\lambda$) then
controls brane-world effects on the tensor CMB anisotropies in
this simple local approximation.

Note that the 4D metric perturbation variable, $H_T$, which
characterizes the amplitude of 4D gravitational waves, is related
in flat models to the covariant variables by
\begin{equation} \label{e:HT1}
H_{Tk} = \frac{\sigma'_k}{k} + 2 E_k \, .
\end{equation}
If the photon anisotropic stress $\pi_k$ can be neglected,
Eq.~(\ref{e:sigmadot}) implies
 \be
k H_{Tk} = - \sigma_k' - (\zeta + 2){\cal H} \sigma_k\,.
 \ee

\section{ CMB Tensor Power Spectra}
%%%%%%%%%%%%%%%%%%%%%%%%%%%%%%%%%%%

In the tight coupling regime we can neglect the photon anisotropic
stress (i.e. $\pi_{k}=0$), and the variable
\begin{equation}
u_k \equiv a^{1+{\zeta}/{2}} \sigma_k
\end{equation}
satisfies the equation of motion
\begin{equation} \label{e:ueqn2}
u_k'' + \left[k^2 +2 K - \frac{(a^{-1-
{\zeta}/{2}})''}{a^{-1-{\zeta}/{2}}} \right] u_k = 0\, ,
\end{equation}
by Eq.~(\ref{app}). In flat models ($K=0$) on large scales, the
solution is
\begin{equation}
\sigma_k= A_k a^{-(2+\zeta)} + B_k a^{-(2+\zeta)} \int d\tau\,
a(\tau)^{2+\zeta} \,,
\end{equation}
where $A_k$ (decaying mode) and $B_k$ are constants of
integration. If we let $\zeta \to 0$, we recover the general
relativity solution.

We can solve Eq.~(\ref{e:ueqn2}) on all scales in the high-energy
($\rho\gg \lambda$ and $a\propto \tau^{1/3}$) and low-energy
($\rho\ll\lambda$ and $a\propto \tau$) radiation-dominated
regimes, and during matter-domination ($a\propto \tau^2$). The
solutions are
\begin{eqnarray}
\label{e:highenergy1} u_k(\tau) &=&\sqrt{k\tau} \left[ c_1
J_{(5+\zeta)/6} (k \tau)  + c_2 Y_{(5+\zeta)/6}
(k \tau)\right] ~~(\mbox{high energy radiation}), \\
\label{e:lowenergy1} u_k(\tau) &=&\sqrt{k\tau} \left[c_3 J_{
(3+\zeta)/2} (k \tau) + c_4 Y_{(3+\zeta)/2}
(k \tau)\right] ~~(\mbox{low energy radiation}),\\
\label{e:matter} u_k(\tau) &=&\sqrt{k\tau} \left[ c_5
J_{(5+2\zeta)/2} (k \tau)  + c_6 Y_{(5+2\zeta)/2} (k \tau)\right]
~~(\mbox{matter domination}),
\end{eqnarray}
where $c_i$ are integration constants and $J_n, Y_n$ are Bessel
functions. The solutions for the electric part of the brane Weyl
tensor can be found from Eq.~(\ref{e:sigmadot}). For modes of
cosmological interest the wavelength is well outside the Hubble
radius at the transition from the high energy regime to the low
energy. It follows that the regular solution (labelled by $c_1$)
in the high-energy regime will only excite the regular solution
($c_3$) in the low-energy, radiation-dominated era. Performing a
series expansion, we arrive at the appropriate initial conditions
for large-scale modes in the low-energy radiation era:
\begin{eqnarray}
\label{e:hijini} H_{Tk} &=& 1 - \frac{(k \tau)^2}{2 (3 + \zeta)} +
\frac{(k \tau)^4}{8 (3+\zeta)(5+\zeta)}  + O[(k\tau)^6], \\
\sigma_k &=&-\frac{k \tau}{3+\zeta} + \frac{k^3
\tau^3}{2(3+\zeta)(5+\zeta)} + O[(k\tau)^5], \label{e:shearini}\\
\label{e:elecini} E_k &=& \frac{(4 + \zeta)}{2(3+\zeta)} -
\frac{(k\tau)^2 (8+\zeta)}{4(3+\zeta)(5+ \zeta)} + O[(k\tau)^4].
\end{eqnarray}
In the limit $\zeta \to 0$, we recover the general relativity
results~\cite{c}.

For modes that are super-Hubble at matter-radiation equality
(i.e.\ $k \tau_{\rm eq} \ll 1$), the above solution joins smoothly
onto the regular solution labelled by $c_5$ in
Eq.~(\ref{e:matter}). For $k \tau_{\rm eq} \gg 1$, the shear
during matter domination takes the form
\begin{equation}
\label{e:matterlong} \sigma_k = -
2^{(3+2\zeta)/2}\Gamma[(5+2\zeta)/2] (k\tau)^{-(3+2\zeta)/2}
J_{(5+2\zeta)/2}(k\tau).
\end{equation}
In the opposite limit, the wavelength is well inside the Hubble
radius at matter-radiation equality. The asymptotic form of the
shear in matter domination is then
\begin{equation}
\label{e:mattershort} \sigma_k \sim
\frac{\Gamma[(3+\zeta)/2]}{\sqrt{\pi}} \left(
\frac{2\tau_{\text{eq}}}{\tau}\right)^{1+\zeta/2}
(k\tau)^{-(1+\zeta/2)} \sin\left(k\tau - {\pi\over 4}\,\zeta
\right).
\end{equation}

\begin{figure}[!bth]
\begin{center}
\includegraphics[scale=0.6]{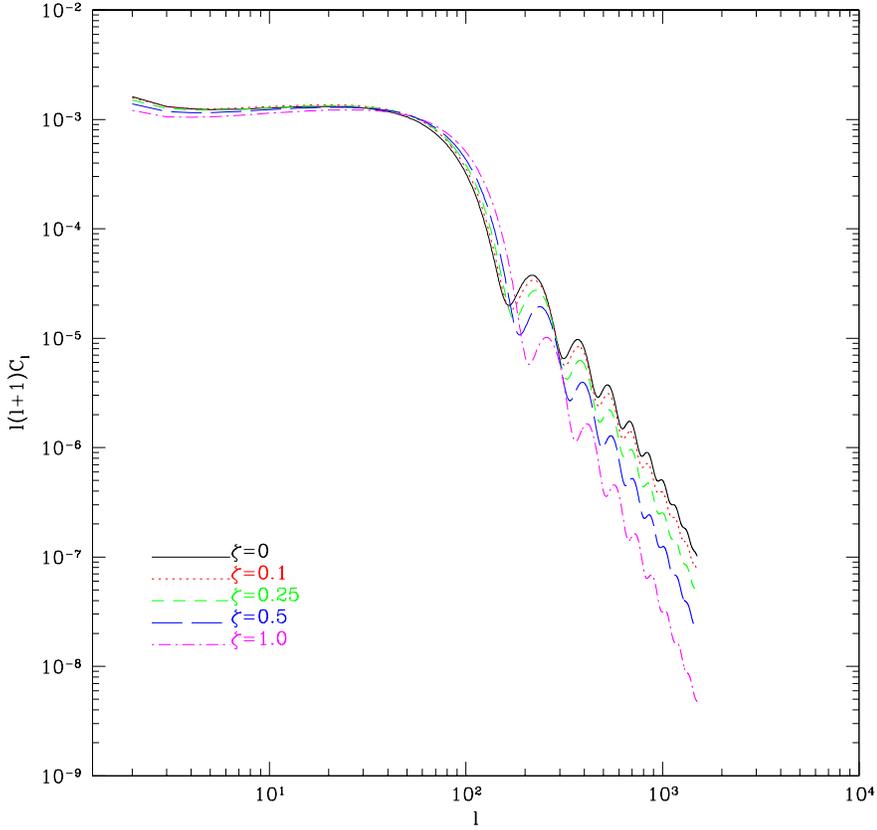}[angle=-90]
\caption{The temperature power spectrum for tensor perturbations
in brane-world models, using the approximation in
Eq.~(\ref{e:Pansatz1}), with $\zeta$ the dimensionless KK
parameter. Models are shown with $\zeta= 0.0$,  0.1, 0.25,  0.5
and 1.0. The initial tensor power spectrum is scale invariant. The
background cosmology is the (concordance) spatially flat
$\Lambda$CDM model with density parameters $\Omega_b=0.035$,
$\Omega_{c}=0.315$, $\Omega_{\Lambda}=0.65$, no massive neutrinos,
and the Hubble constant $H_0=65\, {\rm kms}^{-1} {\rm Mpc}^{-1}$.
 }
\end{center}
\end{figure}

The initial conditions Eqs.~(\ref{e:hijini})--(\ref{e:elecini}),
are used in a modified version of the CAMB code~\cite{camb} to
obtain the tensor temperature and polarization power
spectra~\cite{l2}. The temperature and electric and magnetic
polarization spectra are shown in Figs.~2--4 for a scale-invariant
initial power spectrum. The normalization is set by the initial
power in the gravity wave background. Within the local
approximation to $\cp_k$, the power spectra are insensitive to
high-energy effects: the $\zeta=0$ curve in Fig.~2 is
indistinguishable from that of the general relativity model. For
the computations, the lowest value of the brane tension $\lambda$,
consistent with the limit $\lambda
>(100~{\rm GeV})^4$, is used, but the results are largely
insensitive to the value of $\lambda$ within the local
approximation.

There are three notable effects arising from our approximation to
the KK stress:\\ (1)~the power on large scales reduces with
increasing KK parameter $\zeta$;\\ (2)~features in the spectrum
shift to smaller angular scales with increasing $\zeta$; \\
(3)~the power falls off more rapidly on small scales as $\zeta$
increases.

Neglecting scattering effects, the shear is the only source of
linear tensor anisotropies~\cite{c}. For $1\ll l < 60$ the
dominant modes to contribute to the temperature $C_l$s are those
whose wavelengths subtend an angle $\sim 1 / l$ when the shear
first peaks (around the time of Hubble crossing). The small
suppression in the $C_l$s on large scales with increasing $\zeta$
arises from the reduction in the peak amplitude of the shear at
Hubble entry [see Eq.~(\ref{e:matterlong})], qualitatively
interpreted as the loss of energy in the 4D graviton modes to 5D
KK modes.

Increasing $\zeta$ also has the effect of adding a small positive
phase shift to the oscillations in the shear on sub-Hubble scales,
as shown e.g.\ by Eq.~(\ref{e:mattershort}). The delay in the time
at which the shear first peaks leads to a small increase in the
maximum $l$ for which $l(l+1)C_l$ is approximately constant, as is
apparent in Fig.~2. The phase shift of the subsequent peaks in the
shear has the effect of shifting the peaks in the tensor $C_l$s to
the right. For $l > 60$ the main contribution to the tensor
anisotropies at a given scale is localized near last scattering
and comes from modes with wavenumber $k \sim l / \tau_0$, where
$\tau_0$ is the present conformal time. On these scales the
gravity waves have already entered the Hubble radius at last
scattering. Such modes are undergoing adiabatic damping by the
expansion and this results in the sharp decrease in the
anisotropies on small scales. Increasing the KK parameter $\zeta$
effectively produces more damping and hence a sharper fall off of
power. The transition to a slower fall off in the $C_l$s at $l
\sim 200$ is due to the weaker dependence of the shear amplitude
on wave-number at last scattering for modes that have entered the
Hubble radius during radiation domination. The asymptotic
expansion of Eq.~(\ref{e:matterlong}) gives the shear amplitude
$\propto k^{-(2 + \zeta)}$ at fixed $\tau$, whereas for modes that
were sub-Hubble at matter-radiation equality
Eq.~(\ref{e:mattershort}) gives the amplitude $\propto k^{-(1 +
\zeta/2)}$.

\begin{figure}[!bth]
\begin{center}
\includegraphics[scale=0.6]{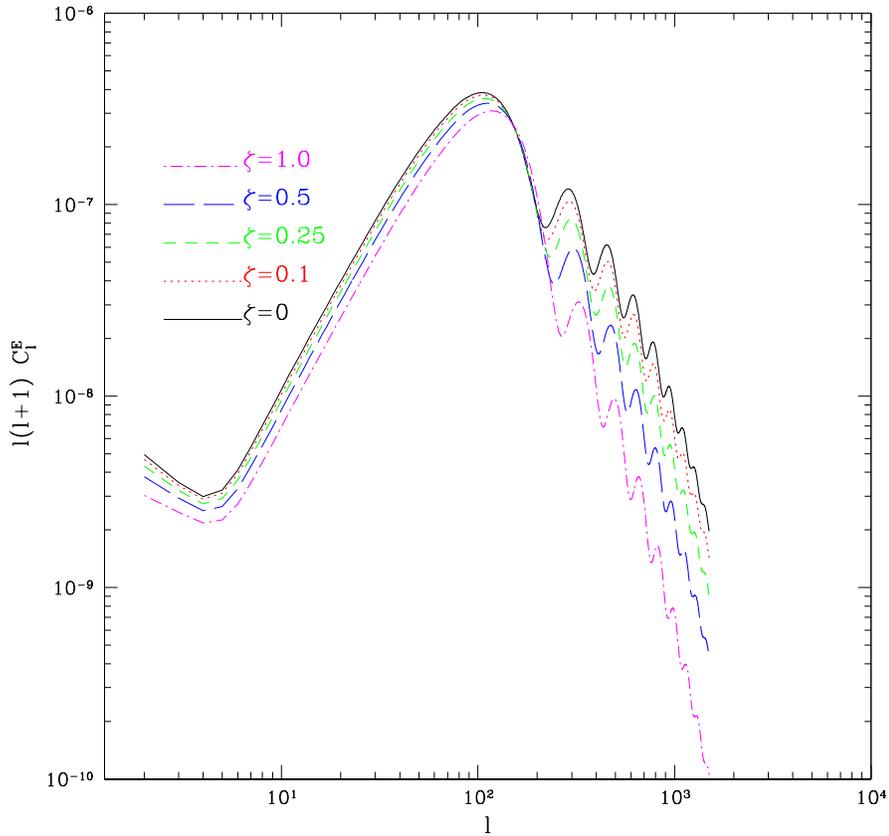}[angle=-90]
\caption{ The electric polarization power spectrum for tensor
perturbations for the same brane-world models as in Fig.~2.
 }
\end{center}
\end{figure}

Similar comments apply to the tensor polarization $C^E_l$ and
$C^B_l$, shown in Figs.~3 and 4. As with the temperature
anisotropies, there is the same shifting of features to the right
and increase in damping on small scales. Since polarization is
only generated at last scattering (except for the feature at very
low $l$ that arises from scattering at reionization, optical depth
0.03) the large-scale polarization is suppressed, since the shear
(and hence the temperature quadrupole at last scattering) is small
for super-Hubble modes. In matter domination the large-scale shear
is $\sigma_k = - k\tau / (5 + 2\zeta)$; the reduction in the
magnitude of the shear with increasing KK parameter $\zeta$ is
clearly visible in the large-angle polarization.

\begin{figure}[!bth]
\begin{center}
\includegraphics[scale=0.6]{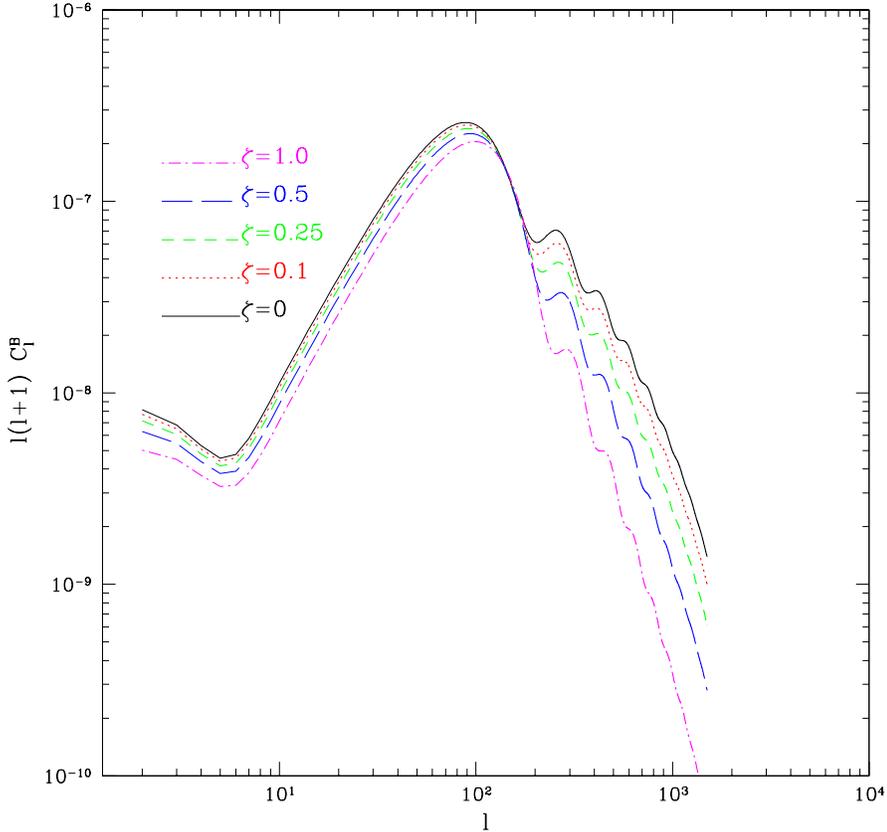}[angle=-90]
\caption{ The magnetic polarization power spectrum for tensor
perturbations for the same brane-world models as in Fig.~2.
 }
\end{center}
\end{figure}

\section{Conclusion}
%%%%%%%%%%%%%%%%%%%%

In principle, observations can constrain the KK parameter $\zeta$,
which determines the brane-world effect on tensor anisotropies in
the CMB by controlling the generation of 5D modes within a
simplified local approximation, Eq.~(\ref{e:Pansatz1}). The other
brane-world parameter $\lambda$, the brane tension, is not
constrained within this approximation. This may indicate that the
approximation entails a hidden low-energy assumption, or it may
simply be accidental.

The local approximation to $\cp_k$ introduced in~\cite{l2} is a
first step towards the calculation of the brane-world imprint on
CMB anisotropies on small scales, starting with the simplest case
of tensor anisotropies. In practice, the tensor power spectra are
not measured, and the prospect of useful data is still far off.
What is more important is the theoretical task of improving on the
simplified local approximation in Eq.~(\ref{e:Pansatz1}). This
approximation encodes aspects of the qualitative features of
brane-world tensor anisotropies, primarily the loss of energy in
4D graviton modes via 5D graviton emission, which may be expected
to survive in modified form within more realistic approximations.
However, a proper understanding of brane-world effects must
incorporate the nonlocal nature of the KK graviton modes, as
reflected in the general form of Eq.~(\ref{e:soln}).

Furthermore, it is the scalar anisotropies which dominate the
measured power spectra, and it is therefore of even greater
importance to develop the scalar analysis. In this case, even the
first step of a local approximation has to confront the problem of
the frame ambiguity in Eq.~(\ref{e:Pansatz1}). An alternative
frame-invariant local approximation is needed, as a first step.
Nonlocal approximations, and ultimately numerical integration of
the nonlocal equation for $\cp_k$, must be developed.

Once these more realistic nonlocal approximations are developed
and the brane-world imprint on the CMB is computed, there are two
key tasks:

\begin{itemize}

\item
generalize the CMB brane-world computations to include a bulk
scalar field and other extensions of the RSII-type models;

\item
determine the implications of the brane-world imprint for the
fundamental theory, M~theory, which underlies the key aspects of
phenomenological brane-world cosmologies.

\end{itemize}

\[ \]
{\bf Acknowledgments}

I am grateful to the organisers of the workshop at the Yukawa
Institute for support and warm hospitality. I thank Anthony
Challinor, Bernard Leong, Kei-ichi Maeda, Shinji Mukohyama and
Takahiro Tanaka for very helpful discussions, and Bernard Leong
for Fig.~4. I acknowledge support from PPARC.

\end{document}